\documentclass[reprint,
amsmath,amssymb,
aps,
]{revtex4-2}

\usepackage{graphicx}% Include figure files
\usepackage{dcolumn}% Align table columns on decimal point
\usepackage{bm}% bold math
\usepackage{hyperref}

\usepackage{amsmath,amssymb,amsthm}
\usepackage[margin=1in]{geometry}
  
\newcommand{\ket}[1]{\lvert#1\rangle}
\begin{document}

\preprint{APS/123-QED}

\title{Noise-Tolerant Object Detection and Ranging Using Quantum Correlations}% Force line breaks with \\

\author{Hashir Kuniyil}
 
\author{Helin Ozel}%
\affiliation{%
Dept. of Electrical and Electronics Engineering, Özyeğin University, Istanbul, Turkey, 34794
}%
\author{Kadir Durak}%
 \email{kadirac@gmail.com}
\affiliation{%
 Dept. of Electrical and Electronics Engineering, Özyeğin University, Istanbul, Turkey, 34794 
}%

\author{Hasan Y\i lmaz}
 
\affiliation{
 Institute of Materials Science and Nanotechnology, National Nanotechnology Research Center (UNAM), Bilkent University, 06800 Ankara, Turkey
}%

\date{\today}% It is always \today, today,

\begin{abstract}
Imaging, detection and ranging of objects in the presence of significant background noise is a fundamental challenge in optical sensing. Overcoming the limitations imposed in conventional methods, quantum light sources show higher resistance against noise in a time-correlation-based quantum illumination. Here, we introduce the advantage of using not only time correlations but also polarization correlations in photon pairs in the detection of an object that is embedded in a noisy background. In this direction, a time- and polarization-correlated photon pair source using the spontaneous parametric down-conversion process is exploited. We found that the joint measurement of correlated pairs allows distinguishing the signal from the noise photons and that leads to an improved signal-to-noise ratio. Our comparative study revealed that using polarization correlations in addition to time correlations provides improved noise rejection. Furthermore, we show that polarization correlation allows undoing the detector limitation where high background often leads to detector saturation.
\end{abstract}

\maketitle

Detection of an object which is embedded in a noisy background is a fundamental problem in sensing and imaging. Conventional techniques explore the properties of the electromagnetic field to reconstruct the structure of an object under investigation. This type of detection scheme is limited by the classical bounds~\cite{berchera2019quantum}. These limitations are fundamental in the classical realm, and therefore overcoming its limitations requires non-classical detection methods~\cite{dowling2003quantum}. Non-classical correlation measurement has been recognized as a candidate to surpass classical limits offering advancements in many technological frontiers such as quantum cryptography~\cite{ekert1991quantum,acin2006bell}, quantum computing~\cite{steane1998quantum, preskill2018quantum}, quantum imaging~\cite{aspden2015photon, kolenderska2020quantum}, quantum ranging~\cite{zhang2015entanglement, luong2019receiver}, and quantum microscopy~\cite{white1998interaction}. We are specifically looking into the scope of quantum correlations in detection and ranging of an object under a noisy background, that is of interest in LIDAR applications.

A quantum-entangled source-based detection scheme called quantum illumination (QI) enables an exponential improvement in the signal-to-noise ratio (SNR) of detecting a sample compared to a classical source-based detection within a noisy environment~\cite{lloyd2008enhanced}. QI requires a phase-sensitive joint measurement scheme that utilizes the spontaneous parametric down-conversion process (SPDC)~\cite{klyshko1988photons} to generate photon pairs that have non-classical spatial and temporal correlations~\cite{lopaeva2013experimental}. However, the practical implementation of QI is restricted with limitations imposed by the losses in the spatial correlations after scattering by diffusive surfaces in realistic scenarios. Thus, this system is only appropriate to use for an object that has an optically smooth surface. To overcome these problems, only non-classical temporal correlation of the pairs are often employed in QI detection scheme~\cite{england2019quantum, liu2020target, kuniyil2020object, durak2019object}. Although phase-insensitive scheme does not completely offer the QI's advantages such as simple experimental settings, high response time compared to classical systems~\cite{liu2020target}, and possibility to tune the wavelength of the source~\cite{paterova2020quantum, aspden2015photon} (especially for quantum imaging applications), the quantum phase-insensitive scheme is still preferable over the classical ones. In phase-insensitive systems, a test for the value of the second order correlation function ($g^{(2)}$) can be used to determine if the source is operating in the quantum or classical regime~\cite{england2019quantum}. If the $g^{(2)}$ value is above the classical regime~\cite{tan2021quantum} of 2, it is a direct indication that the source is operating in the quantum regime and the correlation of the source is higher than any classical temporal correlation.

In this letter, we use SPDC-generated pairs, correlated both in temporal and polarization degrees of freedom, as a quantum source. A configurable noise is externally injected to test the system's behaviour in the presence of noise. A test for obtaining $g^{(2)}$ values is carried out by considering only temporal correlation (TC) of the source and compared to the case of both temporal and polarization correlations (TPC) in identical settings. Our results indicate that TPC has higher second order coherence compared to TC in a noisy environment; therefore, improved resistance can be achieved by using both temporal and polarization correlations.

\begin{figure}[htbp]
\centering
\includegraphics[width=\linewidth]{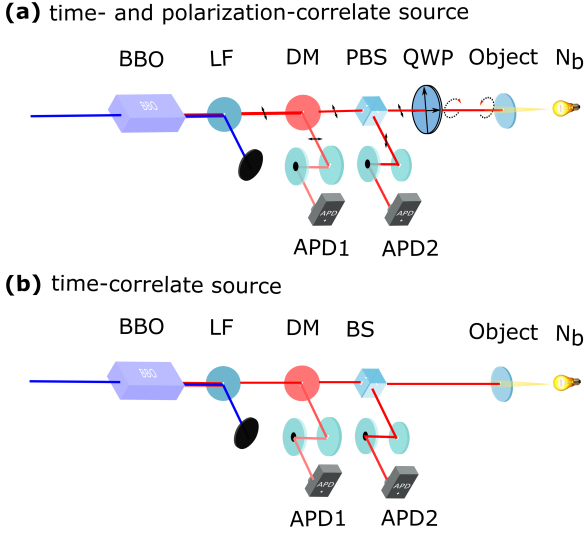}
\caption{The experimental arrangement used for testing the detection schemes that uses (a) time and polarization correlations and (b) only time correlations.}
\label{fig:false-color}
\end{figure}

A schematic of the experimental setup used for this study is shown in Fig.~\ref{fig:false-color}. The correlated photon pairs generated in a SPDC process are used for the quantum illumination. Conventionally, these pairs are called signal and idler photons. For the ease of explanation, we call "the signal photons" for photons that are scattered by the target and "reference photons" for the idler photons recorded locally. A continuous-wave (CW) narrow bandwidth laser of 405 nm wavelength and 160 MHz linewidth is used to pump  a second-order nonlinear medium ($\chi^{(2)}$), generating a co-polarized signal and reference photons in the SPDC process. The twin photons are generated by the SPDC process by utilizing a type-1 BBO crystal where the signal and the reference have central wavelengths of 842 nm and 780 nm, respectively. The blue pump is blocked after the BBO crystal using a long-pass filter with a cut-off wavelength of 750 nm. Signal and reference photons are also separated by a dichroic mirror with a cut-off wavelength of 805 nm. The reference photons are directly measured with the first avalanche photo-detector (APD1) while the signal photons pass through a polarizing beam splitter (PBS) and a quarter-wave plate (QWP) before it scatters back over the object (see in Fig.~\ref{fig:false-color}a). The horizontally polarized signal photons become right-hand circularly (RHC) polarized at QWP in the forward path and returns by left-hand circularly (LHC) polarized after being scattered back by the object. The back scattering from the object happens at the $2\pi$ solid angle but only the light that is scattered back withing the angular range of the illumination optics is collected as the same optics is used both for back and forth paths. In the return path, the LHC signal photons become vertically polarized after the QWP consequently reflects at the PBS and is detected by the APD2. In the time correlated case, which is shown in Fig.~\ref{fig:false-color}b, the signal photons are transmitted from the dichroic mirror and then half of the photons pass through a beam splitter. Half of the back-scattered signal photons is reflected towards the APD at the beam splitter (BS). However, for a fair comparison the source intensity in the TC case is increased to obtain the same number of pair sources for both the TC and the TPC cases. The outputs of two APDs are fed into a timestamp unit (not shown in Fig.~\ref{fig:false-color}) where the timing information of signals is recorded. An unpolarized and broadband external noise ($N_b$) is injected to study the performance of the system under the noisy background. In our settings, the quantum state of polarization of the signal and reference photons are $\ket{H_s}$ and $\ket{H_r}$, where $H$ indicates the horizontal polarization. The '$s$' and '$r$' indices refer to the signal and reference photons, respectively. The polarization of the signal photons is flipped $90^\circ$ in Fig. \ref{fig:false-color}b before detection at APD2, and turn the combined state of polarization into $\ket{V_s}$ and $\ket{H_r}$, where $V$ refers to the vertical polarization. First, we used an anodized aluminium (reflectance is $0.13 \pm 0.03$~\cite{durak2019object}) as a target object to test if the detection is possible and the polarization correlation of the source is maintained after scattering from a rough surface. Therefore, some amount of polarization state will be sacrificed even in specular reflection \cite{nee1996polarization, nee1992ellipsometric}. As a result, the probability of detection of returned signal photons is extremely low. However, we have detected a coincidence detection of $200\pm10$ when the target is positioned closer (10 cm) to the transmitter (distance from the PBS and object in Fig. 1.(a) with the pump power to be 128 $\mu$m. The anodized aluminium is later replaced with a semitransparent mirror as a target object to allow the injection of noise photons externally. An incandescent lamp was used as a power-tunable broadband unpolarized thermal source to inject noise. Using a broadband source devises to replicate a more realistic case where the spectral filtering cannot extinguish the noise. Moreover, such a thermal light source acts as a proper noise model in a correlation-based detection scheme since it shows high classical correlations~\cite{lopaeva2013experimental}. The noise is unpolarized, which makes it a difficult scenario to filter out the noise from signal not only in the presence of only temporal correlations between the signal and reference photons, but even in the presence of polarization correlations. Here, the temporal correlation refers to the fact that both the signal and reference photons are generated almost simultaneously.
\begin{figure}[htbp]
\centering\includegraphics[width=\linewidth]{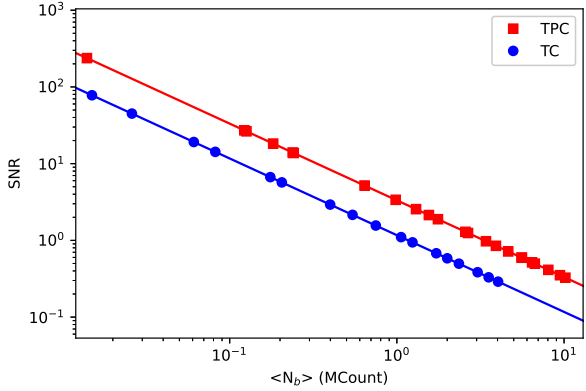}
\caption{The measured signal to noise ratio (SNR) versus the injected noise $N_b$ is plotted for time correlation (TC) (blue dots) and time and polarization correlation (TPC) (red dots) schemes. The theoretical fit (solid lines) is made by using equation 1.}
\label{fig:comparison}
\end{figure}
In the experimental setup we use only phase-insensitive quantum correlation, where the signal arm has extra delay with respect to the reference arm. The timestamp unit is configured to search for joint arrival of photons from both APDs. If the signal and reference photons statistically coincide at a certain delay between them it appears on the cross-correlogram. The timestamp unit we use has a timing resolution (time bin width) of $81$ ps. Therefore, if two photons arrive within this time interval, that is considered as a coincidence. In our settings, the signal photons travel $1.2$ m extra path, so a coincidence is not expected even though both photon pairs are created together within the 81 ps. To obtain a coincidence in this case, we scan the bin width until an avalanche of coincidence is observed. In our experimental configuration, the maximum coincidence value observed, when the pump power is set to 126, $\mu$m, is $5300\pm100$ pairs per second with noise-free settings. We lose some pairs due to the losses in the system and non-unitary reflection coefficient of the object. These values are set approximately the same for both TC and TPC (Fig. 1) cases to make a reasonable comparison. To quantify the quantum correlation exploited by the experimental setup, we estimate $g^{(2)}$ value from the observed coincidence and singles values of the signal and reference. The suitable definition of $g^{(2)}$ for our type of system is identified to be~\cite{fortsch2015highly,  guo2017parametric, lee2016highly} 
\begin{align}
         g^{(2)}(\tau) = \frac{C(\tau)}{N_sN_r \Delta t T} 
\label{eq:g2}         
\end{align}
    
where $\tau$ is the timing delay between the signal and reference photons, $C$ the observed number of coincidence at a delay of $\tau$, $N_s$ and $N_r$ are the detection events per unit time for the signal and reference photons, respectively. $\Delta t$ is the resolution of the timestamp unit and $T$ is the duration of the experiment. The maximum $g^{(2)}$ value for the case of TC and TPC in a noise-free condition is calculated to be $260\pm 16.1$ and $261\pm16.2$, respectively. By increasing the noise level, a dramatic reduction in the $g^{(2)}$ value for both cases are observed. However, in the TPC case, the reduction is considerably smaller. The classical bound for $g^{(2)}$ is between 1 and 2 with 2 is an indication of the highest classical correlation. 
\begin{figure}[htbp]
\centering\includegraphics[width=\linewidth]{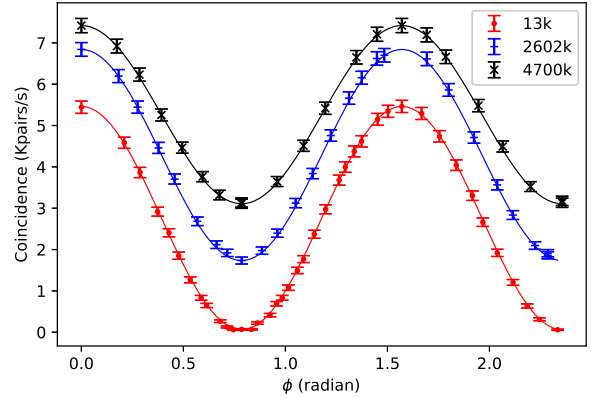} 
\caption{Coincidence value showing a sinusoidal trend while rotating the quarter wave plate, which effectively controls the amount of detected signal that is reflected from the object. This behaviour is observed only from the polarized signal photons (since the noise is unpolarized). The increased level of noise rises the floor height of the sinusoidal plots (minimum coincidence value is not zero). Therefore, floor height gives rise to the estimation of the accidental coincidence. The error bars are obtained by assuming the photon fluctuates by  $\sqrt{N}$. Solid lines represent theoretical fit (see appendix.\ref{abc} for details).}
\label{fig:sin}
\end{figure}

\begin{figure}[htbp]
\centering\includegraphics[width=\linewidth]{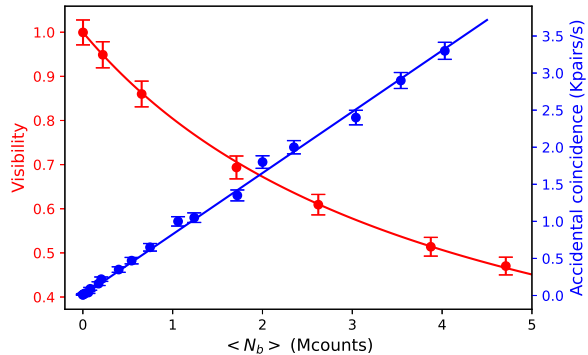} 
\caption{Demonstrating the variation of visibility with respect to the average noise (red). The accidental coincidence increases linearly with the external noise (blue). Points represent the experimental data and the solid lines represent the theoretical fit using the equation (3). Error bars in the accidental coincidence are the order of $\sqrt{N}$. The error in the visibility is calculated by following the standard error propagation method.}
\label{fig:vis}
\end{figure}

The behavior of signal to noise ratio (SNR) with various input noise levels is shown in Fig.~\ref{fig:comparison}. By consistently increasing the amount of injected noise, we estimated the $g^{(2)}$ value. With injected noise, a proportional drop in the SNR values are observed for both TC and TPC settings, with TPC having less decrease compared to TC. This reduction is influenced by the rate of the accidental coincidence imposed by the external noise. The ratio of the SNR in TPC case to TC case is found to be fixed and approximately 2.85. The factor $ N_sN_i\Delta t T$ in equation~\ref{eq:g2} is a direct measure of the accidental coincidence. Accidental coincidences are measured when the signal arm is blocked before the PBS in Fig.~\ref{fig:false-color}a and before the BS in Fig.~\ref{fig:false-color}b by allowing the noise photons to be detected by APD2. Here, note that, $g^{(2)}$ is the total coincidence value including the accidental ones. Therefore, the signal level in the SNR is indicated by $g^{(2)}-1$ to subtract the effect of noise. We observed linear relationship between accidental coincidence value and the injected noise (see Fig.~\ref{fig:vis}). Therefore, a $g^{(2)}$ value of 261 clearly shows the quantum nature of the pairs from the SPDC source.

The polarization correlation between the signal and the reference arm, in addition to the temporal correlations, provides distinguishability in coincidence measurements between the SPDC and the noise photons. Fig.~\ref{fig:sin} shows a sinusoidal behavior of the reflected light from the object due to the polarization correlations between signal and reference photons. The plot shifts upward in accordance with the increasing noise. This is realized by rotating the quarter-wave plate. Combination of PBS and the QWP (see figure (1)) ensures that only $90^\circ$-rotated vertical linearly polarized light is detected with the APD. In this regard, rotating the QWP in the optical axis and evaluating the change in coincidence detection allow differentiating the signal from the noise (hence accidental coincidence). The magnitude of the sinusoidal plots is $C_{max} - C_{min}$, where $C_{max}$ and $C_{min}$ are the measure of coincidence when both signal and reference photons are present and only noise is present cases, respectively. The magnitude reduces as the level of photons detected is reaching the pinch-off region of the detectors. This behaviour is attributed to the APDs' functional limitations when they operate near the saturation region. To be specific, the detectors will not sense correlated photons efficiently due to detector's inability to reset themselves fast enough when more photons are present \cite{grieve2016correcting}.

The control of the amount of signal photons in the TPC system opens up a new way of estimating the presence of noise, leading to the possibility of further noise rejection. In this context, we introduce a figure of merit parameter called visibility that can be defined as 

\begin{equation}
    V = \frac{C_{max} - C_{min}}{C_{max} + C_{min}}
    \label{eq:vis}
\end{equation}
In the absence of background noise, $C_{min}$ becomes zero leading to the visibility value being unity. An estimation of visibility over accidental coincidence is plotted in Fig.\ref{fig:vis}. The relation of the visibility and the accidental coincidence can be estimated by following the relation (see the appendix. \ref{abc} for the derivation):
\begin{equation}
    V = \frac{C_{corr}(N_s)}{C_{corr}(N_s)+2C_{ac}(N_b)d}
    \label{eq:vism}
\end{equation}
where $C_{corr}$ is the coincidences obtained between the signal and the idler, $N_s$ and $N_b$ are the number of the signal and noise photons per unit time, respectively, $C_{ac}$ is the accidental coincidence and $d$ is the detector's correction factor (details of the effect of detector parameters on the visibility is given in the appendix. \ref{abc}). The $d$ parameter is substituted to the equation to address the change in visibility value due to detector's operational change in detection rates close to saturation. The deduced equation fits with experimental data as shown in Fig.~\ref{fig:vis}. The derived expression is suitable and useful for finding the visibility value irrespective of detectors' characteristic limitations. Previous study shows that visibility of the single photon detector is highly dependent on the detector saturation \cite{grieve2016correcting} i.e., if the detectors are operating in near or deep saturation region the actual visibility will be miscalculated. The derived expression for visibility is applicable in any situation. Because the saturation of the detector is guided by the dead time of the detector. In the given expression dead time of the detector is normalized. This calculation will prevent misleading conclusions based on constraints from the detector’s internal electronics circuitry in QI detection schemes.  

In conclusion, we investigated the prospect of using time and polarization correlations simultaneously in quantum illumination for sensing an object that is present in the noisy background. The polarization correlation of the source is often ignored in the consensus that scattering from an object will lead to a complete (or substantial) loss of them. Our study shows that the polarization correlation is maintained even after scattering by a diffusive surface and it improves the SNR compared to the case where only temporal correlations are used in the presence of background noise. Although, the visibility decays with injected noise to the setup, the SNR is improved 2.85 times thanks to the polarization correlations of the pairs. Additionally, polarization correlation of the signal and reference pairs allows to separate the signal from the noise. We quantify that and give a useful expression that also ignores detector parameters such as detector dead time and saturation. This study suggests the interrogation of polarization entanglement preservation via violation of Bell's inequality upon the detection and ranging of an object in quantum illumination scheme.

\section*{Funding} This work was supported by Scientific and Technological Research Council of Turkey (TUBITAK)
with project number 118E991.

\section{appendix}\label{abc}
\subsection{Quarter wave plate rotation versus sinusoidal behaviour on the coincidence}
In a noise-free setting, the orientation of the principal axis (fast axis) of the quarter-wave plate defines the maximum and minimum coincidences. Following Mallus's law, this condition can be expressed as \begin{align}
    C = C_{max}cos(2\theta)
    \label{eq:E1}
\end{align} where C is the number of coincidence defined by the rotation of the quarter-wave plate's axis, $C_{max}$ is the maximum coincidence value between the signal and the reference photons; this occurs when the fast axis of the QWP is at 0, $\pi/2$, $\pi$, $3\pi/2$..etc to linearly polarized signal photons. Therefore, the coincidence maximum is observed when $\theta = 0^\circ$ and the coincidence minimum at $\theta = 45^\circ$. Intensity of the external noise is unaffected by the quarter-wave plate's reference axis ($\theta$) as the noise is a broadband unpolarized light. This means that the rotation of the quarter-wave plate will change only the coincidence between signal and idler (not accidental coincidence).
\subsection{Detector saturation effect on visibility}
The visibility (a contrast parameter) of the polarization correlated coincidence detection scheme is defined by 

\begin{align}
    V = \frac{C_{max} - C_{min}}{C_{max} + C_{min}}
    \label{eq:E2}
\end{align}

where $C_{max}$ and $C_{min}$ are the measure of coincidence when both signal and reference photons are present and only noise is present cases, respectively. In a practical settings, the $C_{max}$ is the summation of the coincidence between the signal and reference and the accidental coincidence. Therefore, the maximum observed coincidence value can be rewritten as $C_{max} = C_{corr}(N_s)+C_{ac}(N_b)$, where $C_{corr}$ is the coincidences obtained between the signal and idler, $N_s$ and $N_b$ are the number of the signal and noise photons, respectively and $C_{ac}$ is the accidental coincidence. The $C_{min}$ can be equate to the accidental coincidence. After incorporating these information, Eq.~\ref{eq:E2} can be rewritten as 

\begin{align}
    V = \frac{C_{corr}(N_s)+C_{ac}(N_b)-C_{ac}(N_b)}{C_{corr}(N_s)+C_{ac}(N_b)+C_{ac}(N_b)}
    \label{eq:E3}
\end{align}
turns into, 
\begin{align}
    V = \frac{C_{corr}(N_s)}{C_{corr}(N_s)+2C_{ac}(N_b)}
    \label{eq:E4}
\end{align}
APD's parameters also play an important role in the visibility of the source. It is because that the APDs do not give actual photon count it encounters above threshold. Figure (1) shows the APD's output (as photon counts in the figure) for the various input photons. The experimental results (doted points) shows a linear trend when less number of photons is sent and it flattens for high photon number values. The solid line (blue) is an estimated plot considering APD behaves same regardless of the input value (called actual count). In the experiment, the intensity of the transmitted signal is constant and the noise is increased to test for system's behaviour in different noise ($N_b$) conditions. Since the accidental coincidence is a function of the noise, defined by $C_{ac} = (N+N_b)N_r \tau$, it also follows the trend of $N_b$. 
\begin{figure}[htbp]
\centering\includegraphics[width=\linewidth]{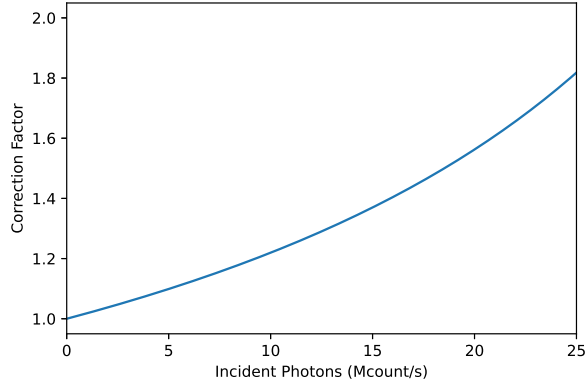}
\caption{Correction factor versus incident photons}
\label{fig:SI}
\end{figure}To incorporate this concern we introduce the detector function, called correction factor, into the Eq.~\ref{eq:E4} and it turn to 
\begin{align}
    V = \frac{C_{corr}(N_s)}{C_{corr}(N_s)+2C_{ac}(N_b)d}
    \label{eq:E5}
\end{align}
where d is the detector correction factor which is defined as $d= 1/(1-AV\times t_d)$, where AV is the actual value observed and $t_d$ is the dead time of the detector. This is consistent with \cite{grieve2016correcting}, and we observe that correction factor is the inverse of the effective duty cycle. We have used M/S. Excelitas make (series no: SPCM-AQRH-TR) actively quenched APD with dead time of 18 ns as a receiving detector that collect the returned signal from the object. We have calibrated the APD for input values and found the detector behaviour for different input photons as shown Fig. ~\ref{fig:SII} The figure shows that if the noise level (input photons in the figure)is high the detector gives a imprecise value. That means if we do not account the detectors characteristics the actual visibility will be miscalculated.

\begin{figure}[htbp]
\centering\includegraphics[width= \linewidth]{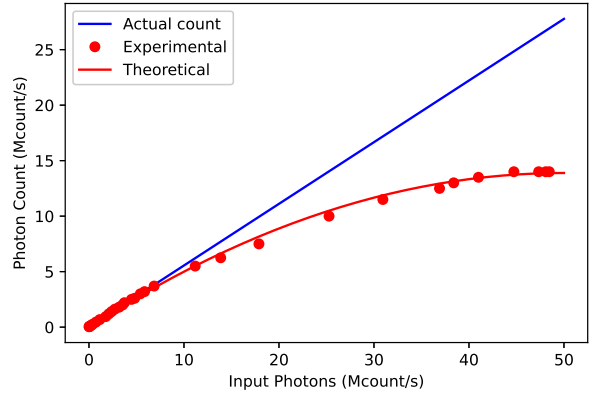} 
\caption{The incident photon versus input photon plot. Red dots are experimentally obtained photon count rate at the detector. A power meter value converted into photon number gives rise to input photons count. }
\label{fig:SII}
\end{figure}
\bibliography{ref}
\end{document}